\documentclass[12pt]{article}
\usepackage{./asp2014}

\aspSuppressVolSlug
\resetcounters

\begin{document}

\title{High-resolution spectrograph for  telescopes of moderate diameter}

\author{V.~Panchuk,$^{1,2}$ M. Yushkin,$^1$ V.~Klochkova,$^1$ Yu.~Verich,$^1$ and G.~Yakopov$^1$ 
\affil{$^1$Special Astrophysical Observatory, Nizhnij Arkhyz, Karachai-Cherkessia, Russia;  \email{panchuk@ya.ru}}
\affil{$^2$ITMO University, S.-Peterburgh, Russia;}}

\paperauthor{V. Panchuk}{panchuk@ya.ru}{ORCID_Or_Blank}{Special Astrophysical Observatory}{Astrospectroscopy laboratory}{Nizhnij Arkhyz}{Karachai-Cherkessia}{369167}{Russia}
\paperauthor{M. Yushkin}{maks@sao.ru}{ORCID_Or_Blank}{Special Astrophysical Observatory}{Astrospectroscopy laboratory}{Nizhnij Arkhyz}{Karachai-Cherkessia}{369167}{Russia}
\paperauthor{V. Klochkova}{valenta@sao.ru}{ORCID_Or_Blank}{Special Astrophysical Observatory}{Astrospectroscopy laboratory}{Nizhnij Arkhyz}{Karachai-Cherkessia}{369167}{Russia}
\paperauthor{Yu. Verich}{yu.verich@gmail.com}{ORCID_Or_Blank}{Special Astrophysical Observatory}{Observations support laboratory}{Nizhnij Arkhyz}{Karachai-Cherkessia}{369167}{Russia}
\paperauthor{G. Yakopov}{yakopov@sao.ru}{ORCID_Or_Blank}{Special Astrophysical Observatory}{BTA technical support team}{Nizhnij Arkhyz}{Karachai-Cherkessia}{369167}{Russia}

\begin{abstract}
A base model of the high-resolution fiber--fed spectrograph is developed. In combination with 
the SAO 1-meter telescope the spectrograph has the following parameters: spectral resolution R\,=\,45000, 
the number of simultaneous registered orders is 86 within spectral region 3850$\div$10850\,\AA{}, 
echelle orders are overlapping for $\lambda <9000$\,\AA{}. 
\end{abstract}

\section{Description of the spectrograph}

A base model of the high-resolution fiber-fed spectrograph for moderate diameter telescopes is developed. 
Its stationary  part (Fig.\,\ref{fig1}) is performed under the white pupil layout which was proposed
by~\citep{Baran}. 

\articlefigure[width=1.2\textwidth]{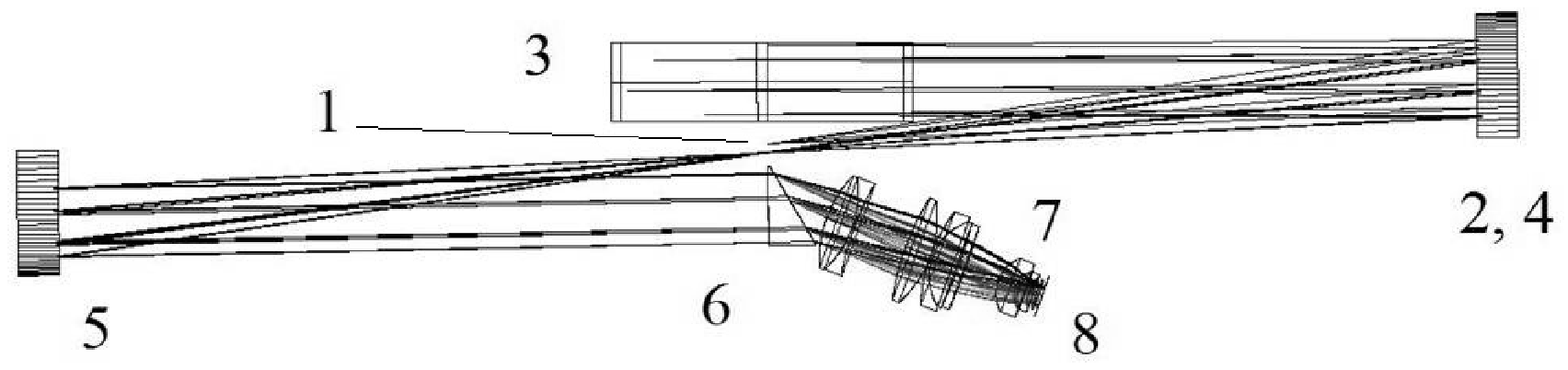}{fig1}{Stationary part of the spectrograph. Legend: 1 -- fiber input, 
               2 -- collimator,  3 -- echelle, 4  -- the first collector who generates an intermediate 
               spectrum,  5 -- the second collector, 6 -- cross-disperser grism, 7 -- lens camera, 8 -- CCD.}

A moving part contains a polarimetric unit (Fig.\,\ref{fig2}), a guiding system and three fibers.

There were two schemes of the observations: spectropolarimetric and Doppler, each only uses two optical fibers. 
In the spectropolarimetic mode register simultaneously two spectra with different circular polarization, 
in the Doppler's mode  are being registered simultaneously stellar  and Th-Ar lamp spectra.
The comparison spectrum contains bright lines (see Fig.\,\ref{fig3}), therefore simultaneous registration 
of star and lamp spectra will distort both of them.

\articlefigure[width=0.8\textwidth]{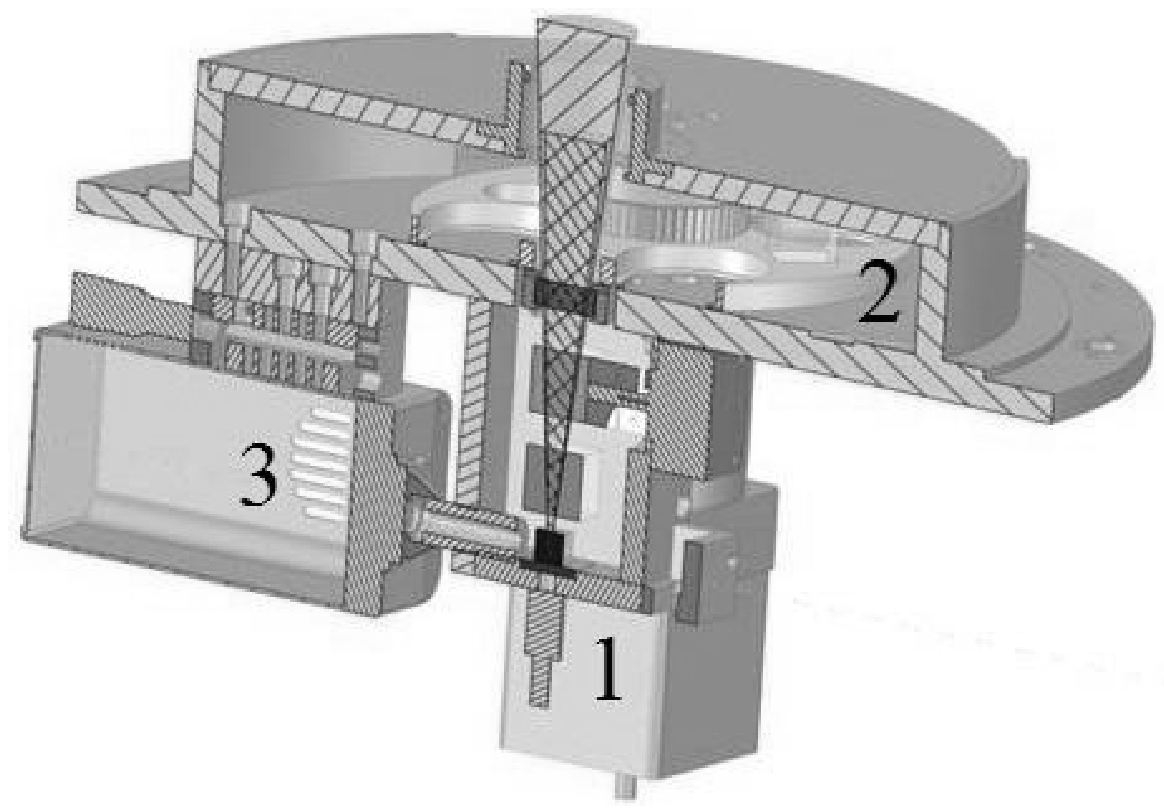}{fig2}{The layout of the moving part of the spectrograph.  
                   Notes:   1 -- fiber inputs,   2 -- polarimetric turret,  3 -- TV-camera of the autoguiding system.}  

\articlefigure[width=0.8\textwidth]{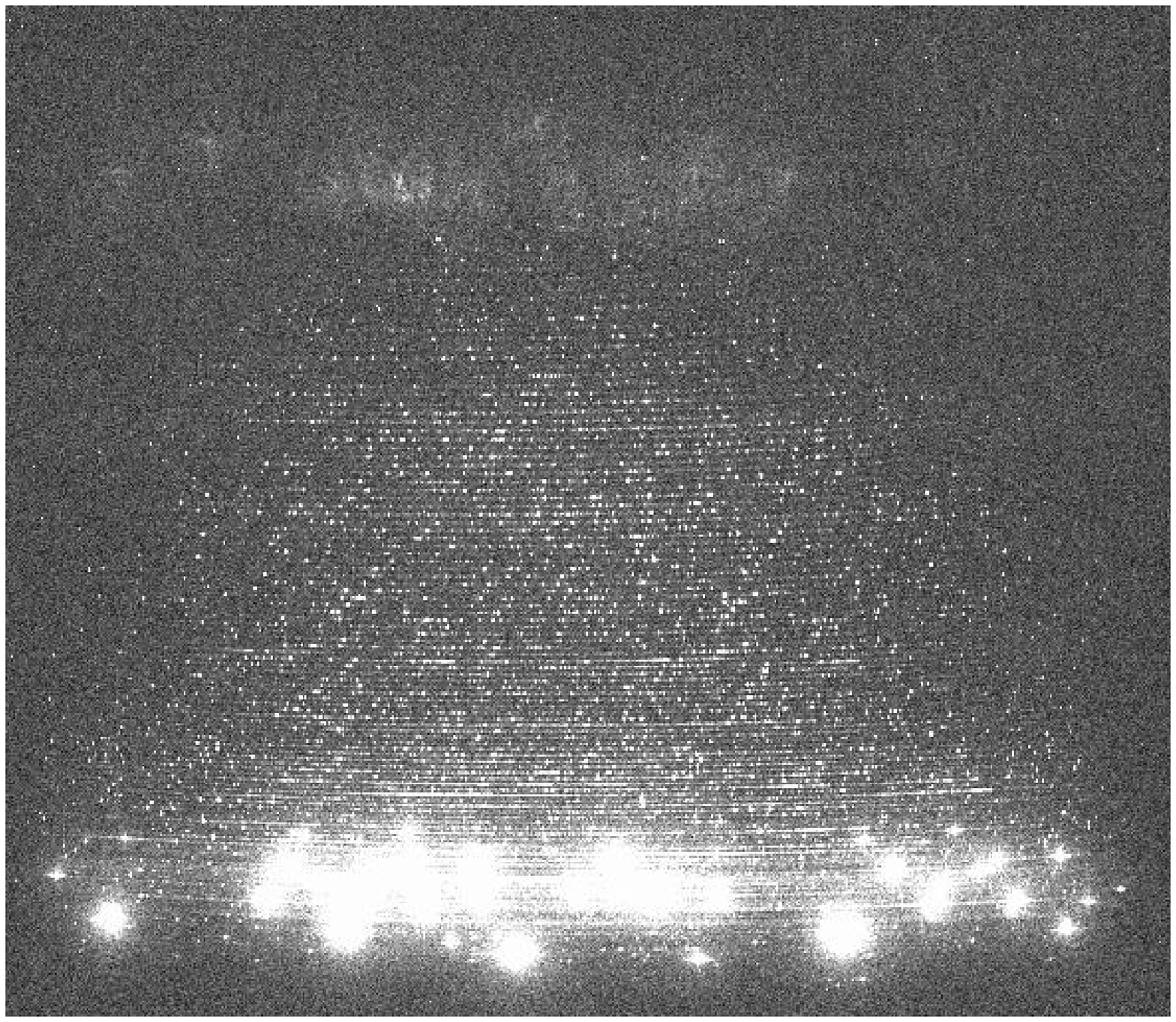}{fig3}{Spectrum of the Th-Ar lamp registered with the spectrograph 
             ARGES~\citep{ARGES}}
                                   
A special device for removing these distortions because of the bright lines and to regulate the flux from the 
Th-Ar lamp was developed. This device provides parallel and simultaneous registration of spectra of star and 
Th-Ar lamp with exposures of reasonable duration.

The control system of the spectrograph is based on the National Instruments technology and contains 
the controller NI CompactRIO 9073  which  includes five modules. Programm code is written within the environment 
of the graphics programming LabView.

In combination with the SAO 1--m telescope the spectrograph provides the following parameters: 
spectral resolution R\,=\,45000, the number of simultaneously registered orders is 86 within the spectral 
region  3850$\div$10850\,\AA{} (moreover, each echelle order is registered twice).  
On the CCD 2048$\times$2048 pixels, echelle orders are being overlapped for $\lambda <9000$\,\AA{}.

The spectrograph is designed for observations of objects whose regular monitoring at the BTA is 
not possible. The image of the spectrum  of the heavily reddened star  (V\,=\,11.7\,mag, 
B\,=\,14.0\,mag) obtained when testing budget version of the spectrograph~\citep{Panchuk} is presented in 
Fig.\,\ref{fig4}.

\articlefigure[width=0.8\textwidth]{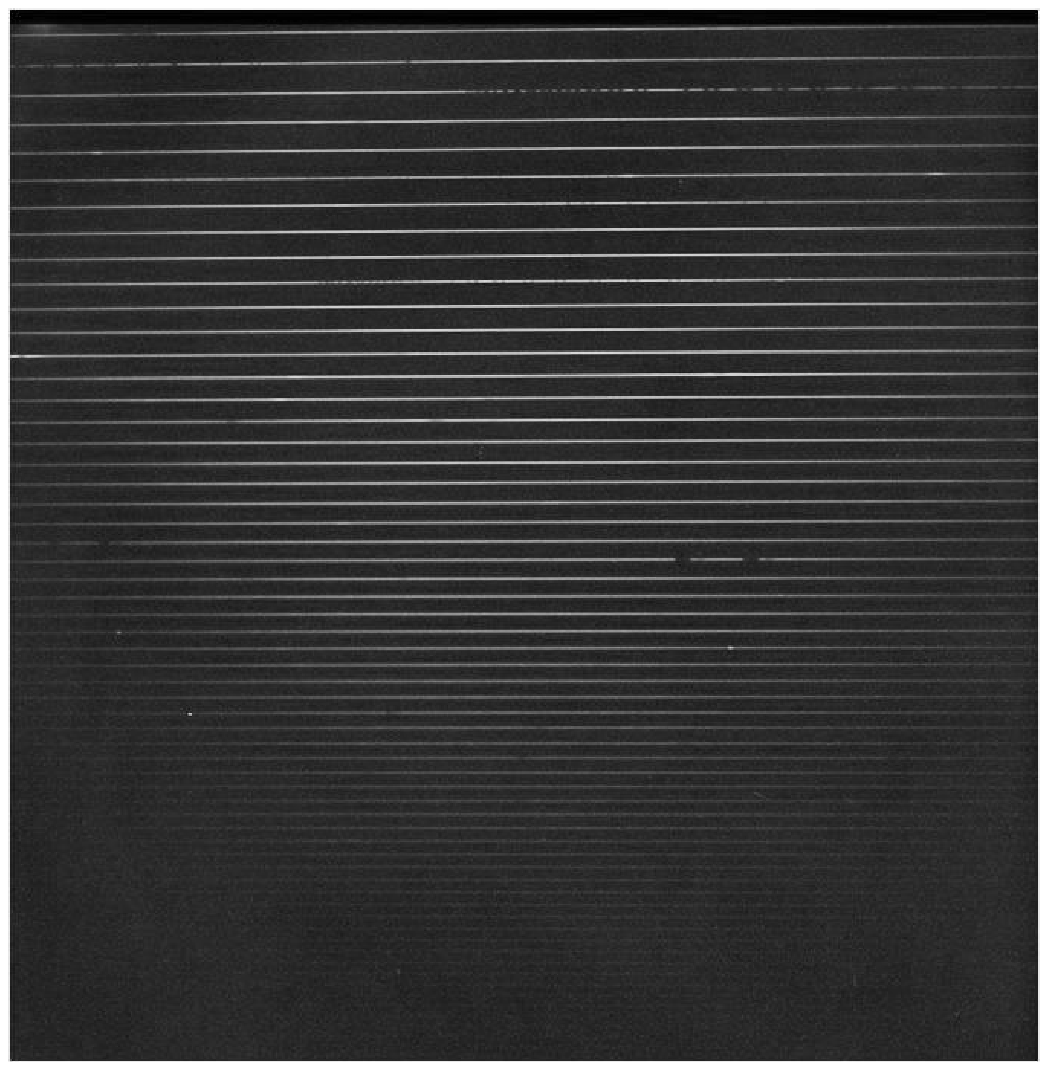}{fig4}{Spectrum of V1302\,Aql (V\,=\,11.7\,mag, B\,=\,14.0\,mag) 
     registered with the spectrograph at the 1--meter  SAO telescope. Long wavelength orders are top.}

\acknowledgements 
We acknowledge financial support by the Russian Foundation for Basic Research (projects 12--07--00739, 
13--02--00029, 14--02--00291). 


\end{document}